\documentclass[prodmode,acmtap]{ci2018}

\acmVolume{2}
\acmNumber{3}
\acmArticle{1}
\articleSeq{1}
\acmYear{2010}
\acmMonth{5}

\usepackage[ruled]{algorithm2e}
\SetAlFnt{\algofont}
\SetAlCapFnt{\algofont}
\SetAlCapNameFnt{\algofont}
\SetAlCapHSkip{0pt}
\IncMargin{-\parindent}
\renewcommand{\algorithmcfname}{ALGORITHM}

\title{Optimizing the Human-Machine Partnership with Zooniverse}
\author{LUCY FORTSON and DARRYL WRIGHT \affil{University of Minnesota}
CHRIS LINTOTT \affil{University of Oxford}
LAURA TROUILLE \affil{Adler Planetarium}
}

\begin{abstract}
Intentionally left blank.
\end{abstract}

\begin{document}

\maketitle

\section{Introduction: Citizen Science as a Human-Machine Partnership }

Over the past decade, Citizen Science has become a proven method of distributed data analysis, enabling research teams from diverse domains to solve problems involving large quantities of data with complexity levels which require human pattern recognition capabilities. With over 120 projects built reaching nearly 1.7 million volunteers, the Zooniverse.org platform has led the way in the application of Citizen Science as a method for closing the Big Data analysis gap \cite{Watson2018}.  
Since the launch in 2007 of the Galaxy Zoo project \cite{Lintott2008,Fortson2012}, the Zooniverse platform has enabled significant contributions across many disciplines; e.g., in ecology \cite{Swanson2016},  humanities \cite{Williams2014a}, and astronomy \cite{Marshall2015}. 
Citizen science as an approach to Big Data combines the twin advantages of the ability to scale analysis to the size of modern datasets with the ability of humans to make serendipitous discoveries \cite{Marshall2015}.

To cope with the larger datasets looming on the horizon such as astronomy's Large Synoptic Survey Telescope (LSST) or the 100's of TB from ecology projects annually,  
Zooniverse has been researching a system design that is optimized for efficiency in task assignment and incorporating human and machine classifiers into the classification engine.
A growing body of theoretical work, often using data from Zooniverse projects, has demonstrated that efficiencies exist in task assignment to volunteers that could greatly reduce the burden on classifiers (e.g.\, \cite{Simpson2013,Waterhouse2013,Kamar2015}). 
Furthermore, task assignment studies within the Space Warps project demonstrated that false negatives (fields wrongly classified as containing no lenses) could be eliminated if at least one user with high measured skill reviews each subject \cite{Marshall2016}. 
Other work suggests that efficiencies can be gained through the judicious coupling of machine and human classifiers (e.g.\, \cite{Russakovsky2015,Kamar2012}). For example, presenting tasks in order of increasing machine confidence reduces the time to obtain a given target accuracy by 63\% \cite{Veit2015}.

We hypothesize that by making efficient use of smart task assignment and the combination of human and machine classifiers, we can achieve greater accuracy and flexibility than has been possible to date. We note that creating the most efficient system must take into account that humans are complicated \cite{Mugar2014,Bowyer2015}  and the system must consider how best to engage and retain volunteers as well as make the most efficient use of their classifications. 
Our work thus focuses on understanding the factors that optimize efficiency of the combined human-machine system. This paper summarizes some of our research to date on integration of
machine learning with Zooniverse, while also describing the new infrastructure developed on the Zooniverse platform to carry out this research.

\section{Machine Learning integration with Zooniverse}

Remarkable progress has been made in recent years by machine learning researchers, though at the expense of requiring extremely large training sets. For some  problems 
such training sets can be assembled from simulations, however the vast majority of classification problems face hurdles in building up a large training set; a particular difficulty when searching for rare objects. 
Until now, the frontier for machine learning in citizen science was for researchers to use  \emph{post facto} the large training sets produced by citizen science projects \cite{Beaumont2014,Dieleman2015}.  
New Zooniverse infrastructure described below enables a more sophisticated approach; combining human and machine classifications and optimizing effort via smart task allocation. 

\subsection{The Zooniverse Decision Engine}
Zooniverse projects are supported by the Panoptes codebase \footnote{https://github.com/zooniverse/Panoptes} which is an application of modular design supporting a powerful API. On request, the API serves subjects - images, video or sound files - for classification by volunteers via a workflow defined by the project owners. The core tasks of the Panoptes API are allowing the creation and management of projects, passing subjects to the frontend software for classification, and the receiving and recording of classifications. Further modules handle secondary tasks such as data aggregation, discussion, or the provision of statistics on project progress or user behavior. 
The key functionality that Zooniverse is developing for combining human and machine classifiers rests on backend modules that together decide which subjects from a project's subject pool should be retired from classification (`\emph{Caesar}')  and designate the next round of subject sets to be shown to specific volunteer groups on specific workflows associated with a project (`\emph{Designator}'). 
\emph{Caesar} monitors classifications in real time and is able to perform actions through several distinct modes: `extractors', where relevant features are extracted from a new classification; `reducers', where multiple of these relevant features are processed so that they can be used to form a consensus or decision; `rule application', during which simple comparators can be used to provide rudimentary flow control; and `effects', which act on the Panoptes API to change the state of users, subjects, etc. 
The subject upload manifest incorporates metadata which can include, for example, pre-trained machine scores for any subject; these metadata are associated with the subject throughout the classification process and thus can be used by the decision engine as needed. Extractors and reducers can be set up as external algorithms enabling research teams to incorporate specific models into the decision engine.  

\subsection{Experiments}
Integrating machine learning into a Zooniverse project can take a number of forms.  At its simplest, a pre-trained model can be used to filter subjects \textit{offline} as is the case for the Supernova Hunters project.  Here subjects for which the model predicts a low confidence of being a real supernova candidate are automatically rejected and the remaining subjects uploaded for at least ten volunteers to review.  
Equally, an offline model's confidence can be combined with the aggregated volunteer votes for each subject, which has been shown to improve classification performance \cite{Wright2017}. 

The new \emph{Caesar} infrastructure allows for richer interactions between citizen scientists and machine learning models.  For example, the Camera CATalogue project has trained a model offline for species identification on images that required labeling by five volunteers 
before being retired. To more efficiently classify new data the model predicts a species for each image, then using \emph{Caesar}'s advanced rules, the system considers an image classified if the first two volunteers agree with the model's prediction, reducing human effort by 43\% while maintaining overall accuracy \cite{Willi2018}.

Going a step further, \emph{Caesar}'s  live classifications stream and ability to assign subjects to workflows on the fly allows data flow decisions to be made in near real-time. These functionalities are key for active learning. In contrast to the typical approach of uploading every subject for classification, instead we select only those that are expected to be most informative for a model, speeding up convergence and reducing volunteer effort. The efficiency gains from such a system trained in near real-time have been explored in a sequence of simulations and shown to provide at least a factor of eight increase in the classification rate  for the Galaxy Zoo project \cite{Beck2018}.

Access to up-to-date information on both the subjects (e.g., the machine learning confidence) as well as our human classifiers (e.g., the quality of an individual's performance) also allows \emph{Caesar}  to assign subjects to specific groups of users, for example highly correlated classifiers or groups based on experience.  Subjects can be targeted to those we expect to do the best job \cite{Marshall2016} allowing more challenging subjects to be classified by more experienced volunteers and prioritizing the presumably larger number of simpler tasks for the larger pool of less experienced volunteers as exemplified by the Gravity Spy project \cite{Zevin2017}. Together active learning and targeting specific volunteer groups allow machines and humans to each focus on their strengths, reducing the work for humans while improving the quality of training data for the models.

\begin{figure}[tp]
\centering
\includegraphics[width=11cm]{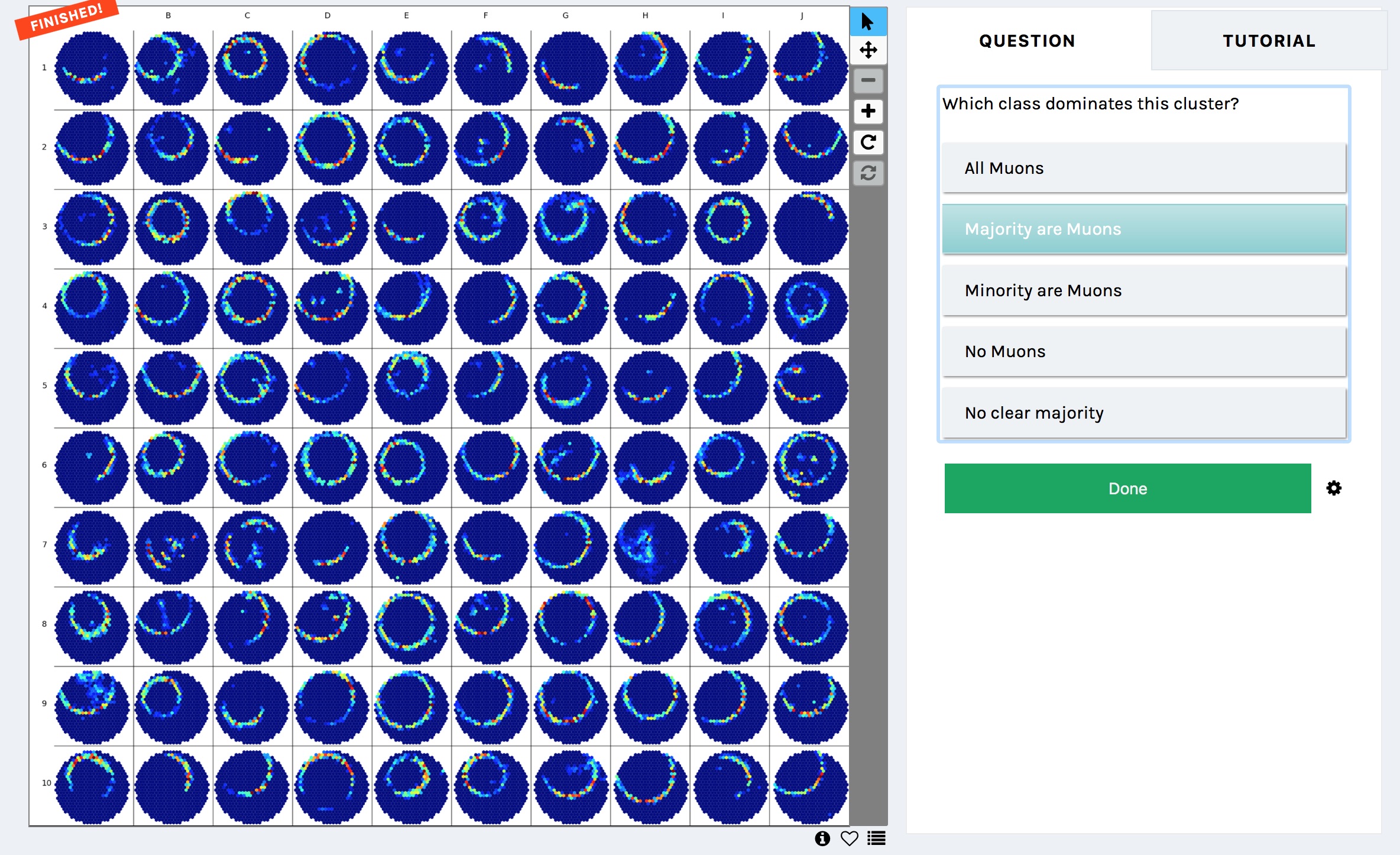}
\caption{Interface from the Zooniverse Muon Hunters project \cite{Bird2018} showing images from unsupervised clustering.}
\label{muonfig}
\end{figure}

Zooniverse hosts a large number of projects that could benefit from machine learning, yet many research groups do not have the expertise to train these models on their volunteer-labeled data. Zooniverse can attempt to build general machine learning tools that,  although not optimized for any specific task, can lead to some of the efficiency gains described above. One example is providing pre-trained models for transfer learning \cite{Willi2018} which can aid smaller camera trap projects.  We have also been experimenting with learning salient feature representations in conjunction with unsupervised clustering to identify meaningful structure in image data uploaded to Zooniverse. Clustering the data provides a different approach to gathering labels from volunteers;  rather than asking volunteers to classify each subject individually, we can instead show similar subjects that have been clustered together. If all subjects in a group belong to the same class citizen scientists can classify the entire group at once, or instead pick out those subjects that look like they may not belong (see Fig.\ref{muonfig}).  Without any knowledge of the task at hand it is easy to surmise that all the images in Fig.\ref{muonfig} belong to the same class because each image is presented in the context of others.  Similarly it is easy to distinguish potential anomalies given this context such as the image in cell H7 of the figure.

In conclusion, Zooniverse has shown the best approach that citizen science can take to optimize knowledge discovery combines both human and machine classiers, and has established critical infrastructure to enable this improved approach to address challenges presented by Big Data.  

\newpage

\bibliographystyle{ci-format}
\bibliography{ci2018-bibfile}

\end{document}